\documentclass[a4paper,10pt,pra,twocolumn]{revtex4}
\usepackage[dvips]{graphics,color}
\usepackage{natbib}
\usepackage{graphicx}
\usepackage{amsmath}
\usepackage{amsfonts}
\newcommand{\one}{\mathrm{I} \! \! 1}
\usepackage{amssymb}
\usepackage{amstext}
\usepackage[sort&compress]{natbib}
\begin{document}

\title{Entanglement and symmetry in permutation symmetric states}

\author{Damian.~J.~H.~Markham} \email{markham@enst.fr}
\affiliation{CNRS, LTCI, Telecom ParisTech,
37/39 rue Dareau, 75014 Paris, France}

\begin{abstract}
We investigate the relationship between multipartite entanglement and symmetry, focusing on permutation symmetric states. We give a highly intuitive geometric interpretation to entanglement via the Majorana representation, where these states correspond to points on a unit sphere. We use this to show how various entanglement properties are determined by the symmetry properties of the states. The geometric measure of entanglement is thus phrased entirely as a geometric optimisation, and a condition for the equivalence of entanglement measures written in terms of point symmetries. Finally we see that different symmetries of the states correspond to different types of entanglement with respect to interconvertibility under stochastic local operations and classical communication (SLOCC).
\end{abstract}
%
\maketitle

\section{Introduction}
Entanglement and symmetry are two main concepts at the heart of quantum mechanics. For a while now there have been enticing hints of a general connection between the two. On an intuitive level, we may understand that changing global symmetry (or topological properties) should be a global operation, so one that effects the entanglement of the system at hand. The relationship is of great interest in particular because of the relation between symmetry and phase transitions. There are by now a vast array of examples where entanglement shows some relationship to symmetry breaking, for example in quantum phase transitions where phase transition coincides with change in entanglement properties \cite{Osterloh02,Osborne02,Vidal03}. Indeed it has been suggested that entanglement may be able to see phase transitions where conventional order parameters fail. However, a concrete relationship remains unclear, for example, it is known change in some symmetry properties need not effect the entanglement, and vice versa. For a recent review of these issues see \cite{Amico08}.

At the same time symmetry properties of states have been used to simplify the study of their entanglement for example in its calculation \cite{Vollbrecht01,Stockton03} and questions of separability \cite{Toth09}. A particular feature of multipartite entanglement is that it is possible to have different `types' of entanglement, where-by we mean different classes under SLOCC (Stochastic Local Operations and Classical Communication) \cite{Dur00}. This property has been largely overlooked in the in the study of phase transitions and the use of symmetries in entanglement theory. Two states are SLOCC inequivalent (belong to different classes) if they cannot be converted to one-another via local operations and classical communications, even probabilistically, which signifies them as potentially different resources in the context of quantum information processing (QIP). Alongside this comes a plethora of entanglement measures, with a variety of different operational interpretations, and which may be suited to quantifying one type of entanglement more than another.
The question naturally arises, can symmetry help us to explore this vast landscape, and can a relationship between symmetry and the types of entanglement be made.

In this work we focus on permutation symmetric states. These states are useful in a variety of QIP tasks, occur naturally as ground states for example in some Hubbard models, and certain of these states have been implemented experimentally recently \cite{Prevedel09,Wieczorek-09}.
Various entanglement properties have also been studied of these states, such as the clarification of separability conditions \cite{Toth09}, the calculation of the geometric measure of entanglement \cite{Hubner09,Hayashi08} and the identification of SLOCC classes \cite{Bastin09}. In all these cases however, permutation symmetry is essentially used as a tool for simplification in calculations. We would like to see if further symmetry properties can be useful, if a deeper connection between symmetry and entanglement properties can be found, and if there may be some insight into the role of entanglement in many body physics. To this extent we observe that symmetries with respect to local operations (rather than permutation) determine several entanglement properties, with intriguing mirrors in spinor Bose-Einstein Condensates (BEC).

In particular, by using the Majorana representation \cite{Majorana}, we see how symmetry allows us to calculate the geometric measure of entanglement \cite{Shimony95} and identify the most entangled state \cite{note}. Then, we show that the existence of certain symmetry guarantees equivalence of three different entanglement measures - the geometric measure of entanglement, the logarithmic robustness of entanglement \cite{Robustness} and the relative entropy of entanglement \cite{Vedral98}. Finally we will see how the different symmetries reflect different types of entanglement, (in terms of SLOCC classes) indicating an intriguing relationship between symmetries and types of entanglement. We will close with some remarks on occasions these same symmetries coincide with different phases for spinor BEC, and how these states may be generated experimentally.

\section{Entanglement in the Majorana representation}

We first present the Majorana representation \cite{Majorana}. This way of seeing states has been used recently to simplify the classification of symmetric states into SLOCC classes \cite{Bastin09,Mathonet09}. For $n$ qubits, all permutation symmetric states can be written in the form \cite{Majorana}
\begin{eqnarray} \label{Eqn: Symm state}
|\psi\rangle =\frac{e^{i\alpha}}{\sqrt{K}} \sum_{PERM} |\eta_1\rangle |\eta_2\rangle...|\eta_n\rangle,
\end{eqnarray}
where the sum is over all permutations and $K$ is a normalisation constant. The Majorana representation consists of $n$ points corresponding to the $n$ states from this decomposition $|\eta_i\rangle = cos(\theta_i/2)|0\rangle + e^{i\phi_i}sin(\theta_i/2)|1\rangle$ via the standard
Bloch sphere - i.e. a point on the unit sphere at position $\theta_i,\phi_i$. We call these the Majorana Points (MPs), and they define the state up to global phase $e^{i\alpha}$ (see Fig.~\ref{FIG: example}). For more details see the Appendix \ref{Sec: Appendix}.

\begin{figure}[t]
{\resizebox{!}{3cm}{\includegraphics{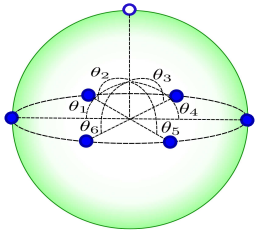}}} \caption{\label{FIG: example} (Colour online) The Majorana representation of the $n$-party GHZ state $|GHZ_n\rangle := (|0\rangle^{\otimes n}+|1\rangle^{\otimes n})/\sqrt{2})$, has $n$ MPs equally spaced around the equator, here for $n=6$ in solid points. The hollow point at the north pole is point of the closest product state, maximizing $\prod_i |\langle \phi|\eta_i\rangle|^2 = \prod_i(cos(\theta_i/2))^2$.}
\end{figure}

To see how entanglement can be visualised in the Majorana representation,  we first notice that the product of local unitaries on a symmetric state $U\otimes U \otimes ... U |\psi\rangle$ is just a rotation of the Majorana sphere, since each point gets rotated by the same $U$. In fact it can be shown that if two symmetric states $|\psi\rangle, |\phi\rangle$ are related by local unitaries $U_1\otimes U_2 \otimes ... U_n |\psi\rangle=|\phi\rangle$, there is always some $U$ such that they can be connected by the same unitary $U\otimes U \otimes ... U |\psi\rangle=|\phi\rangle$. This fact is shown for the more general case of local invertible operations in \cite{Bastin09,Mathonet09}, and the same proof works for unitaries. Further, this shows that the symmetry of the state under local unitaries $U^{\otimes n}$ is reflected by the symmetry of the MPs (see also the Appendix \ref{Sec: Appendix}). This will be a main tool throughout the paper.

We can make the connection to entanglement more explicit, using the {\it geometric measure of entanglement}
\cite{Shimony95},
\begin{eqnarray} \label{Eqn: Def Eg}
E_g(|\psi\rangle) = \min_{|\Phi\rangle \in {\rm PROD}} -\log_2 (|\langle \Phi | \psi \rangle |^2),
\end{eqnarray}
where PROD is the set of product states. It has recently been shown that for permutation symmetric states, we can always take a symmetric product state $|\Phi\rangle= |\phi\rangle |\phi\rangle  ...|\phi\rangle$ in this optimisation \cite{Hubner09}, for which the Majorana representation consists of $n$ points at the position of $|\phi\rangle$. For the
general $n$ partite symmetric state (\ref{Eqn: Symm state}) we then have
\begin{eqnarray} \label{Eqn: Geom Ent of Symm}
E_g(|\psi\rangle) &=& -\log_2 (\Lambda (\psi)), \nonumber \\
\Lambda(\psi) &=& \max_{|\phi\rangle} |\langle \phi|^{\otimes n} |\psi \rangle |^2 \nonumber \\
&=&  \frac{1}{K} n!^2 \max_{|\phi\rangle} |\langle \phi|\eta_1\rangle|^2 |\langle \phi|\eta_2\rangle|^2
...|\langle \phi|\eta_n\rangle|^2. \nonumber
\end{eqnarray}
Hence, the optimisation problem of finding the closest product state has the geometric
interpretation of maximising the product of angles $|\langle \phi|\eta_i\rangle|^2$. The example of $|GHZ_6\rangle$ is illustrated in Fig.~\ref{FIG: example}.

This geometric phrasing of the problem allows us to use geometric properties, for example symmetry of the MP distribution to calculate entanglement and to search for the most entangled states in this class. In a sense, we can say that the most entangled states will be those which spread the points out the most (though this does not necessarily coincide with other definitions of `spread' such as Tammes' problem). This direction is pursued in detail in follow up work \cite{Martin}, and has been independantly studied in \cite{Martin10}.

\begin{figure}[t]
{\resizebox{!}{3cm}{\includegraphics{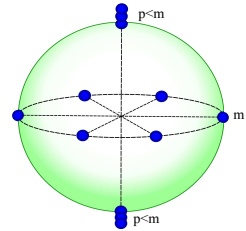}}} \caption{\label{FIG: dihedral} (Colour online) States of $n=m+2p$ qubits which are totally invariant for the Dihedral symmetry groups $D_m$, $|D_m(n,p)\rangle = 1/\sqrt{2} (|S(n,p)\rangle+|S(n,n-p)\rangle)$. The state has $p$ MPs at each pole and $m=n-2p$ MPs distributed evenly around the equator. For all $n,p$, these states satisfy $E=E_{Rob}=E_R=E_G$. }
\end{figure}

\section{Equivalence of entanglement measures}
We will now proceed to see how the Majorana representation can allow us to identify symmetries indicating states for which several distance like entanglement measures coincide, and show that these states represent different SLOCC classes of entangled states.

In ref. \cite{Hayashi08} the relationship between the geometric measure of entanglement and two other distance like entanglement measures, the relative entropy of entanglement $E_R$ \cite{Vedral98}, the logarithmic robustness of entanglement \cite{Robustness} is studied.
In particular, it was shown that if there exists a local unitary group for which the state in question $\psi = |\psi\rangle\langle\psi|$ is itself an invariant subspace of the group, then we have $E_G(|\psi\rangle)=E_R(|\psi\rangle)=E_{Rob}(|\psi\rangle)$.

Equivalence of measures is desirable for several reasons. Primarily because the different measures have different interpretations. For example, $E_{Rob}$ signifies the ability of the state to withstand noise \cite{Robustness} and the relative entropy being an entropic quantity $E_R$ naturally has several information theoretic interpretations \cite{Vedral98}. Since $E_G$ is easier to calculate, it is easier to verify these operational properties also. Further significance of the equivalence is discussed in \cite{Hayashi08} in particular it significance for local accessibility of information and the construction of optimal entanglement witnesses.

Using the equivalence between symmetry of points and of states, we are able to phrase the problem soley in terms of the Majorana representation (see the Appendix \ref{Sec: Appendix Equiv} for more details)
\\

\noindent {\bf Lemma 1}:
\textit{If a permutation symmetric state $|\psi\rangle$ has MPs such that they are invariant under some subgroup $X\subset SO(3)$, and that for any small change of the points this invariance disappears, it satisfies}
\begin{align}
E_G(|\psi\rangle)=E_R(|\psi\rangle)=E_{Rob}(|\psi\rangle). \nonumber
\end{align}
\\

We call such subgroups $X\subset SO(3)$ the symmetry groups, and say such states are totally invariant. Intriguingly, this is exactly the condition for finding inert states in the context of spinor condensates \cite{Makela07}, which will be discussed more in the concluding remarks.

\begin{figure}[t]
{\resizebox{!}{3cm}{\includegraphics{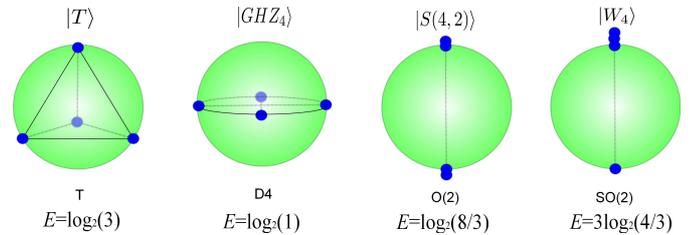}}} \caption{\label{FIG: test} (Colour online) Different symmetries for four qubits giving states such that$E=E_{Rob}=E_R=E_G$. The symmetry group and the entanglement are written below the sphere. Each state is in a different SLOCC class.}
\end{figure}

The Majorana representation then allows us to identify symmetries to show equality of the entanglement measures for many new sets of states. The complete set of all the possible subgroups of SO(3) are the continuous groups, orthogonal $O(2)$ and special orthogonal $SO(2)$, and discrete groups Cyclic $C_m$, Dihedral $D_m$, Tetrahedral $T$, Octehedral $O$ and Isocahedral $Y$. One can then systematically go through all of these groups to find these special states, as done in \cite{Makela07} in the context of inert states. For the subgroup of arbitrary rotations about a fixed axis SO(2), we see that states with MPs only at either pole of the rotation axis satisfy our condition. If the rotations are around the $Z$-axis, these are the states
\begin{align}
|S(n,k)\rangle := \frac{1}{\sqrt{{n\choose k}}}(\sum_{PERM}|\underbrace{00...0}_{n-k}\underbrace{11..1}_{k}\rangle), \nonumber
 \end{align}
also known as Dicke states, and we can see here pictorially the proof of equivalence for these states reported in \cite{Hayashi08}. Note that for even $n$ and $k=n/2$, these states also satisfy our condition for the group $O(2)$ (arbitrary rotations around the $Z$-axis, and a flip on some axis in the $X-Y$ plane). In such cases we associate the state with the smallest subgroup. The cyclic group $C_n$ has no truly invariant states - since if all points are moved together up and down the axis of rotation the symmetry is not lost. The dihedral group $D_m$ (consisting of rotations through $2\pi/m$ and a flip on the axis of rotation) has $m$ totally invariant states for each value $m$ (see Fig.~\ref{FIG: dihedral}). $T$, $O$ and $Y$ only have truly invariant states for certain $n$. For the tetrahedral group $T$ truly invariant states are made up of tetrahedrons, their antipode tetrahedrons, and octagons with at most $2$ MPs on any tetrahedron point and $3$ MPs at any octahedron point, so that there are only truly invariant states for $n\leq 34$. For the octahhedral group $O$ truly invariant states have MPs at the points of the cube and the octahedron with at most $3$ and $2$ MPs at each respectively, so that they only exist for $n\leq34$. All truly invariant states of the Isocahedral group $Y$ are made up of combinations of isocahedrons (with $12$ vertices) and dodecahedrons (with $20$ vertices) with at most $3$ and $2$ MPs at each respectively, hence they exist only for $n\leq88$.  For four qubits there are four entangled states satisfying the condition, $|T\rangle=1/\sqrt{3}|S(4,0)\rangle+\sqrt{2/3}|S(4,3)\rangle$, $|GHZ_4\rangle$, $|S(4,2)\rangle$ and $|W_4\rangle=|S(4,1)\rangle$ as shown in Fig.~\ref{FIG: test}.

\section{Symmetry and SLOCC entanglement classes}
We now look at how these different symmetries also correspond to different SLOCC entanglement classes. First of all, it is shown in \cite{Bastin09,Mathonet09} that if two states have different degeneracies of MPs (that is, the number of MPs which are on top of each other is different), they are SLOCC inequivalent. From this it is clear that:
\\

\noindent {\bf Lemma 2}:
\textit{For any number of qubits greater than two the totally invariant states with respect to the groups $O(2)$, $SO(2)$ and $D_m$ are of different entanglement types.}
\\

This is true since they have different degenerecies. This fact also means that in addition the totally invariant states for the dihedral group $|D_m(n,p)\rangle$ are SLOCC inequivalent for all $p>0$ (see Fig.~\ref{FIG: dihedral}).

For the remaining symmetry groups $T$, $O$ and $Y$ there are only a finite number of possible totally invariant states. We can then use a combination of the degeneracy and other methods to attempt to show the same for these all subgroups. Consider the four qubit case in Fig.~\ref{FIG: test}. From the above, we can see that $|S(4,2)\rangle$ (with two sets of two degenerate MPs) is in a different class to $|W_4\rangle$ (with a three degenerate point), and they are both in different classes to $|T\rangle, |GHZ_4\rangle$ (which have all four MPs separate). To see that the $|T\rangle, |GHZ_4\rangle$ are different, we use the fact \cite{Dur00} that under SLOCC the minimal number of terms $r$ for any expansion of the state in terms of only product states (the log of which is the Schmidt measure \cite{Eisert01}) remains unchanged. It is straightforward to see that taking some minimum decomposition, from definition (\ref{Eqn: Def Eg}) we have $E_G \leq \log_2(r)$. We know that $r(|GHZ_4\rangle)=2$ \cite{Dur00}, and in \cite{Martin} it is shown that $E_G(|T\rangle)=\log_2(3)$, hence $r(|T\rangle)\geq 3 >r(|GHZ_4\rangle)$ and so they are in different SLOCC classes also. For larger $n$ one can in principle go through all cases individually (since there are only finitely many) and check using similar methods. Such an exhaustive search was beyond the scope of this manuscript, however, the same techniques as above can be used to show the SLOCC inequivalence for \textit{all} the totally invariant states up to seven qubits, and it seems plausible that indeed all different symmetries do imply different class of entanglement.

\section{Discussions}

In this work we have presented a geometrical representation of the entanglement of permutation symmetric states in the form of the Majorana representation. This has allowed us to phrase the geometric measure of entanglement in a simple way, and further to look at how the further symmetry properties of states effect their entanglement properties, in particular showing equivalence of three differen distance like measures. This equivalence simplifies calculation, and allows for wider operational understanding as the operational interpretations coincide. Finally we show that for these states the different symmetries correspond to different types of entanglement.
This presents a very interesting relationship between symmetries and types of entanglement. Though we are not able to confirm this is a general connection, it is very interesting, and seems possible, and worth invesgtigating deeper.

Intriguingly, in the context of spinor condensates similar symmetry arguments have been used to identify and characterise different phases of matter \cite{Barnett06,Barnett07,Barnett08}. In this case the Majorana representation is used not to describe $n$ symmetric qubits, but rather a single spin $S=n/2$ system (through the well known isometry between the two) \cite{Majorana}. Because of this caution is required when talking about entanglement in this context, but it is not impossible for the two pictures to coincide, for example, the total spin can be a result of combined spin half systems in exactly the permutation symmetric space we look at, which really would be entangled as we discuss here. In this sense we would see that phase transitions through symmetry are incidental with phase transitions through entanglement, raising the prospect of entanglement type as an indicator of different phases.  Indeed in \cite{Barnett06}, a phase diagram is presented for a spin two spinor condensate where each phase is identified exactly with different symmetry types presented in Fig.~\ref{FIG: test}. Where these connections are most explicit and general is in the case of inert states - often good candidates for ground states in spinor BEC - where the conditions of equivalence of $E_G$, $E_{Rob}$ and $E_R$ coincide exactly in terms of the MPs \cite{Makela07}, pointing to deeper possible connections.

The states discussed here can also be experimentally prepared in a variety of ways and media. For example in optics the six party $|S(6,3)\rangle$ (Dicke) state and several five and four party states have recently been generated, and their entanglement properties verified \cite{Prevedel09,Wieczorek-09}. Further, recently a general scheme has been developed which can generate any symmetric state \cite{Bastin09b} (including all those here) which works for any $\Lambda$-system photon emitters, such as trapped ions or neutral atoms or quantum dots, so may be long lived, and which may be in reach of current experiment.

Since completion of this work several related works have emerged \cite{Martin10,Bastin10,Curt10,Ribeiro11,Aulbach11}.

\noindent {\bf{Acknowledgements}}
We are very grateful to Shashank Virmani, Martin Aulbach, Mio Murao and Vlatko Vedral for insightful comments and discussions.

\appendix

\section{Majorana Representation} \label{Sec: Appendix}


The permutation symmetric subspace of $n$ qubits is spanned by the Dicke states
\begin{align}
|S(n,k)\rangle := \frac{1}{\sqrt{{n\choose k}}}(\sum_{PERM}|\underbrace{00...0}_{n-k}\underbrace{11..1}_{k}\rangle),
\end{align}
which can be understood as the symmetric states with $k$ excitations. Thus any permutation symmetric state can be written as
\begin{align}
|\psi\rangle = \sum a_k |S(n,k)\rangle.
\end{align}
Alternatively all symmetric states can be written in the Majorana representation \cite{Majorana}
\begin{eqnarray} \label{Eqn: App Symm state}
|\psi\rangle =\frac{e^{i\alpha}}{\sqrt{K}} \sum_{PERM} |\eta_1\rangle |\eta_2\rangle...|\eta_n\rangle,
\end{eqnarray}
where the sum is over all permutations and $K$ is a normalisation constant.

To find the Majorana representation (\ref{Eqn: App Symm state}) we consider the overlap with product state $|\phi\rangle^{\otimes n}$,$|\phi\rangle=\cos\left(\frac{\theta}{2}\right)|0\rangle +e^{i \varphi} \sin\left(\frac{\theta}{2}\right)|1\rangle$. It is clear by comparison to equation (\ref{Eqn: App Symm state}) that $|\phi\rangle$ orthoganol to the MP $|\eta_i\rangle$ will give zero overlap. This is exactly how we find the MPs. For simplicity we take a multiple of the overlap, sometimes called the {\it characteristic polynomial}, {\it Majorana polynomial}, {\it amplitude function} or {\it coherent state decomposition}
\begin{align} \label{Eqn: Maj functional}
f(\psi):= \cos^{-n}\left(\frac{\theta}{2}\right) \langle \phi|^{\otimes n}|\psi\rangle = \sum_{k=0}^n \sqrt{{n \choose k}} a_k \alpha^k,
\end{align}
which is a complex polynomial in $\alpha:=e^{-i\varphi}\tan\left(\frac{\theta}{2}\right)$. By the fundamental theorem of algebra this has unique zeros up to multiplication by some complex. Hence the zeros $\alpha_j = e^{-i\varphi_j}\tan\left(\frac{\theta_j}{2}\right)$ define the state $|\psi\rangle$ up to a global phase. The corresponding MPs are at position $\theta_j'=\theta_j+\pi$,~$\varphi_j'=\varphi_j+\pi$.

Note that we can understand the state $|\phi\rangle^{\otimes n}$ as a kind of generalized coherent state \cite{Perelomov,Klauder}, defined by the action of a group on some chosen fiducial state (so that certain properties apply such as overcompleteness). For our case the group is SU(2) as represented by $U^{\otimes n}$, with $U$ a rotation through $\theta, \varphi$ and the fiducial state $|0\rangle^{\otimes n}$, that is
\begin{eqnarray}
|\phi\rangle^{\otimes n}= U^{\otimes n} |0\rangle^{\otimes n}.
\end{eqnarray}
When viewing the symmetric subspace as one spin $S=n/2$ system, these are equivalent to spin coherent states \cite{Arecchi72,Radcliffe71}. In this sense the Majorana representation is a kind of condensed coherent state representation of states (since it is only concerned with the zeros of the coherent state decomposition (\ref{Eqn: Maj functional})).

\bigskip
\section{Equivalence of entanglement measures} \label{Sec: Appendix Equiv}

We now come to the proof of the equivalence between entanglement measures and the symmetry of the MP distributions. The entanglement measures in question are the geometric measure of entanglement \cite{Shimony95}, the relative entropy of entanglement $E_R$ \cite{Vedral98} and the logarithmic robustness of entanglement \cite{Robustness} is studied.
Equality between the measures is guarenteed for a state $|\psi\rangle$, if it is possible to find a separable state of the form \cite{Hayashi08}
\begin{align} \label{Eqn: omega_sep}
\omega_{sep} = \Lambda ( \psi) |\psi\rangle \langle \psi | + (1-\Lambda ( \psi)) \Delta,
\end{align}
where $\Delta$ can be any density matrix and $\Lambda(\psi)$ is the maximum overlap with a product state as defined in (\ref{Eqn: Geom Ent of Symm}). The state (\ref{Eqn: omega_sep}) can be understood as the  `closest' separable state with respect to the robustness of entanglement, which is deemed equal to the geometric measure of entanglement by its form \cite{Hayashi08}.

The trick used in \cite{Hayashi08} is to take techniques from group averaging to find such a state (see also \cite{Vollbrecht01}). For a group $G$, any particular representation $U(g)$, $g\in G$ can always be expanded as a product sum over irreducible representations (irreps, which we enumerate by $k$), and the irreps give a decomposition of the total Hilbert space,
\begin{align}
U(g) &= \bigoplus_{k=1}^K \one_{A_k} \otimes U_{B_k}(g) \\
H &= \bigoplus_{k=1}^K H_{A_k} \otimes H_{B_k},
\end{align}
where $U_{B_k}(g)$ is the representation of $g\in G$ for irrep $k$ acting on Hilbert space $H_{B_k}$. The role of $H_{A_k}$ is just to give a compact form to express multiplicity - the multiplicity of irrep $k$ is given by $dim(H_{A_k}) = Tr(\one_{A_k})$. Note that the tensor product in the above is nothing to do with the separation of parties defining entanglement.
By direct application of Shur's lemma, averaging over the group gives \cite{Hayashi08}
\begin{align} \label{Eqn: omega_twirl}
\omega &= \int U(g) \rho U(g)^\dag dg \nonumber \\
&= \sum_{k=1} \frac{1}{dim(H_{B_k})}Tr_{B_k}\left\{ P_{A_k\otimes B_k} \rho P_{A_k\otimes B_k} \right\} \otimes \one_{B_k}.
\end{align}

If we now average over a local unitary group on a product state $|\Phi\rangle$ which achieves the maximum overlap $\Lambda(\psi)$, we will get a separable state, which is our candidate for (\ref{Eqn: omega_sep}). If, further, the state $\psi = |\psi\rangle\langle\psi|$ is an invariant subspace associated to a one-dimensional irrep (say $k=1$) with multiplicity one we have
\begin{align} \label{Eqn: omega_twirl2}
\omega_{sep} = & \int U(g) \Phi U(g)^\dag dg \nonumber \\
=&\Lambda ( \psi) |\psi\rangle \langle \psi |\nonumber\\
&+ \sum_{k=2} \frac{Tr_{B_k}\left\{ P_{A_k\otimes B_k} |\Phi\rangle \langle \Phi| P_{A_k\otimes B_k} \right\}}{dim(H_{B_k})} \otimes \one_{B_k},
\end{align}
which is indeed of form (\ref{Eqn: omega_sep}), implying equality of the entanglement measures.

In terms of states, $\psi = |\psi\rangle\langle\psi|$ corresponds to a one-dimensional irrep if it is invariant under group action. A one dimensional irrep is a phase which acts over a space of dimension equal to the multiplicity. Any state (one dimensional matrix) in this space is unchanged and so can itself be considered a 1D irrep. Since it is possible to continuously change states through this space, it means that a state which is a 1D irrep, therefore invariant, can be continuously changed to another state which is also a 1D irrep, and hence also invariant. If, on the other hand, a small shift breaks the invariance, the state has multiplicity only one, as we want.

The groups we consider in this work are naturally enough subgroups of $SU(2)$, as represented by the local unitaries $U^{\otimes n}$. Again, we see from the definition (\ref{Eqn: Symm state}), such operations are simply rotations (in SO(3)) of the Majorana sphere itself. Since we are only interested in the state matrix $\psi$ (where global phases do not matter), the invariance the MPs implies a state is a 1D irrep. If no small change in the positions of the points is also invariant, this implies there is no multiplicity within the symmetric subspace. Although this is not immediately enough to show the group averaged state is of the form (\ref{Eqn: omega_sep}), it can be proved as follows. The only remaining possibility for multiplicity is if it has part outside the symmetric subspace. In fact, a projection onto it (say for irrep $k$) must be of the form $P_{A_k\otimes B_k}= |\psi\rangle \langle \psi| + |\psi^\bot\rangle \langle\psi^\bot|$ where $|\psi^\bot\rangle$ has no components in the symmetric subspace. This is true since its representation is $U^{\otimes n}$ and so any 1D irrep cannot stretch over the symmetric subspace and another subspace, but must be distinctly in one or the other. If we put this into (\ref{Eqn: omega_twirl}) (with again $\rho=\Phi$), we indeed get the form (\ref{Eqn: omega_twirl2}).

Thus the condition for equality of measures stated in the main text is correct and complete. For example, for the subgroup of arbitrary rotations about a fixed axis SO(2), we see that states with MPs only at either pole of the rotation axis satisfy our condition. If the rotations are around the $Z$-axis, these are the Dicke states, and we can see here pictorially the proof of equivalence for these states reported in \cite{Hayashi08}.

\bibliographystyle{plainnat}

\begin{thebibliography}{99}


\bibitem{Osterloh02} A.~Osterloh, L.~Amico, G.~Falci and R.~Fazio, Nature
\textbf{416}, 608, (2002).



\bibitem{Osborne02}
T.~Osborne and M.~A.~Nielsen, Phys.~Rev.~A~\textbf{66}, 032110 (2002).


\bibitem{Vidal03}
G.~Vidal, J.~I.~Latorre, E.~Rico and A.~Kitaev, Phys.~Rev.~Lett.~\textbf{90}, 227902 (2003).

\bibitem{Amico08}
L.~Amico, R.~Fazio, A.~Osterloh, V.~Vedral, Rev.~Mod.~Phys.~\textbf{80}, 517-576 (2008).

\bibitem{Vollbrecht01}
K.~G.~H.~Vollbrecht~and~R.~F.~Werner, Phys.~Rev.~A \textbf{64},062307 (2001).

\bibitem{Stockton03}
J.~Stockton, J.~M.~Geremia, A.~Doherty, H.~Mabuchi, Phys. Rev. A \textbf{67}, 022112 (2003).

\bibitem{Toth09}
G.~Toth, O.~G\"{u}hne, Phys.~Rev.~Lett.~\textbf{102}, 170503 (2009).

\bibitem{Dur00}
W.~D\"{u}r, G.~Vidal, J.~I.~Cirac, Phys.~Rev.~A \textbf{62}, 062314 (2000).


\bibitem{Prevedel09}
R.~Prevedel, G.~Cronenberg, M.~S.~Tame, M.~Paternostro, P.~Walther, M.~S.~Kim, A.~Zeilinger, Phys.~Rev.~Lett. \textbf{103}, 020503 (2009).

\bibitem{Wieczorek-09}
W.~Wieczorek, R.~Krischek, N.~Kiesel, P.~Michelberger, G.~Toth, H.~Weinfurter, Phys.~Rev.~Lett. \textbf{103}, 020504 (2009).





\bibitem{Hubner09}
R.~H\"{u}bener, M.~Kleinmann, T.-C.~Wei, C.~Gonzalez-Guillen, O.~G\"{u}hne, Phys.~Rev.~A \textbf{80}, 032324 (2009).

\bibitem{Hayashi08}
M.~Hayashi, D.~Markham, M.~Murao, M.~Owari, S.~Virmani, Phys.~Rev.~A \textbf{77}, 012104 (2008).


\bibitem{Bastin09}
T.~Bastin, S.~Krins, P.~Mathonet, M.~Godefroid, L.~Lamata and E.~Solano, Phys.~Rev.~Lett. \textbf{103}, 070503 (2009).


\bibitem{Majorana}
E.~Majorana, Nuovo~Cimento \textbf{9}, 43-50 (1932).


\bibitem{Shimony95}
A.~Shimony, Ann.~NY.~Acad.~Sci. \textbf{755}, 675 (1995).


\bibitem{note}
Preliminary versions of these results were presented in the QIT 16 workshop in Japan, D.~Markham, Proceedings of QIT 16, Japan (2007).

\bibitem{Robustness} D.~Cavalcanti, Phys.~Rev.~A \textbf{73}, 044302 (2006), G.~Vidal and R.~Tarrach, Phys. Rev. A \textbf{59}, 141 (1999); A.~W.~Harrow and M.~A.~Nielsen, Phys. Rev. A \textbf{68}, 012308
(2003); M.~Steiner, Phys.~Rev.~A \textbf{67}, 054305 (2003).

\bibitem{Vedral98}
V.~Vedral and M.~B.~Plenio, Phys.~Rev.~A \textbf{57}, 1619 (1998).



\bibitem{Mathonet09}
P.~Mathonet, S.~Krins, M.~Godefroid, L.~Lamata, E.~Solano and T.~Bastin , Phys. Rev. A \textbf{81}, 052315 (2010).

\bibitem{Martin}
Aulbach,~M., Markham,~D. and Murao,~M., New J. Phys. \textbf{12} 073025 (2010).



\bibitem{Makela07}
H.~M\"{a}kel\"{a} and K.-A.~Suominen Phys.~Rev.~Lett. \textbf{99}, 190408 (2007).

\bibitem{Eisert01}
J.~Eisert and H.-J.~Briegel, Phys.~Rev.~A \textbf{64}, 022306 (2001).


\bibitem{Barnett06}
R.~Barnett,~A.~Turner~and~E.~Demler, Phys.~Rev.~Lett. \textbf{97}, 180412 (2006).

\bibitem{Barnett07}
R.~Barnett,~A.~Turner~and~E.~Demler, Phys.~Rev.~A \textbf{76}, 013605 (2007).

\bibitem{Barnett08}
R.~Barnett,~S.~Mukerjee and J.~E.~Moore, Phys.~Rev.~Lett. \textbf{100}, 240405 (2008).

\bibitem{Bastin09b}
T.~Bastin, C.~Thiel, J.~von~Zanthier, L.~Lamata, E.~Solano and G.~S.~Agarwal, Phys. Rev. Lett. \textbf{102}, 053601 (2009);
N.~Kiesel, W.~Wieczorek, S.~Krins, T.~Bastin, H.~Weinfurter, E.~Solano, Phys. Rev. A \textbf{81}, 032316 (2010).

\bibitem{Martin10}
J. Martin, O. Giraud, P. A. Braun, D. Braun and T. Bastin, Phys. Rev. A \textbf{81}, 062347 (2010).

\bibitem{Bastin10}
T. Bastin, P. Mathonet, E. Solano, quant-ph/1011.1243 (2010).

\bibitem{Curt10}
C.~D.~Cenci, D.~W.~Lyons and S.~N.~Walck, quantu-ph/1011.5229 (2010).

\bibitem{Ribeiro11}
P.~Ribeiro and R.~Mosseri, quant-ph/1101.2828 (2011).

\bibitem{Aulbach11}
M.~Aulbach, quant-ph/1103.0271 (2011).

\bibitem{Perelomov}
A.~Perelomov, {\it Generalized Coherent States and Their Applications}, Springer-Verlag, Heidelberg (1986).

\bibitem{Klauder}
J.~R.~Klauder and B.~Skagerstam, {\it Coherent States}, World Scientific,~Singapore (1985).

\bibitem{Radcliffe71}
J.~M.~Radcliffe, J.~Phys.~A \textbf{4}, 313-323 (1971).

\bibitem{Arecchi72}
F.~T.~Arecchi and E.~Courtens and R.~Gilmore and H.~Thomas, Phys.~Rev.~A \textbf{6}, 2211 (1972).




\end{thebibliography}

\end{document}